\documentclass{aa}  
\usepackage{graphicx}
\usepackage{color}

\usepackage{txfonts}
\begin{document} 
   \title{{\it Suzaku} observations of the merging galaxy cluster Abell\,2255: \\
The northeast radio relic}
   \author{H. Akamatsu, 
          \inst{1}
          \and
M. Mizuno \inst{2}\and
N. Ota 		\inst{2}\and
Y.-Y. Zhang	\inst{3}\and
R. J. van Weeren \inst{4}\and
H. Kawahara	\inst{5,6}\and \\
Y. Fukazawa	\inst{7}\and
J. S. Kaastra	\inst{1,8}\and
M. Kawaharada  \inst{9}\and
K. Nakazawa	\inst{10}\and
T. Ohashi		\inst{11}\and \\
H.J.A. R\"ottgering \inst{8}\and
M. Takizawa \inst{12}\and 
J. Vink		\inst{13,14} \and
F. Zandanel \inst{14}
          }

   \institute{SRON Netherlands Institute for Space Research, Sorbonnelaan 2, 3584 CA Utrecht, The Netherlands
              \email{h.akamatsu@sron.nl}
         \and
         Department of Physics, Nara Women's University, Kitauoyanishimachi, Nara, Nara 630-8506, Japan
         \and
         Argelander Institute for Astronomy, Bonn University, Auf dem H\"ugel 71, 53121 Bonn, Germany
         \and 
         Harvard-Smithsonian Center for Astrophysics, 60 Garden Street, Cambridge, MA 02138, USA
         \and
         Department of Earth and Planetary Science, The University of Tokyo, Tokyo 113-0033, Japan
         \and 
         Research Center for the Early Universe, School of Science, The University of Tokyo, Tokyo 113-0033, Japan
         \and
         Department of Physical Science, Hiroshima University, 1-3-1 Kagamiyama, Higashi-Hiroshima, Hiroshima 739-8526, Japan
         \and 
         Leiden Observatory, Leiden University, PO Box 9513, 2300 RA Leiden, The Netherlands
         \and
         Department of High Energy Astrophysics, Institute of Space and Astronautical Science, Japan
         \and
         Department of Physics, Graduate School of Science, University of Tokyo, 7-3-1 Hongo, Bunkyo, Tokyo, 113-0033 Japan
         \and
         Department of Physics, Tokyo Metropolitan University, 1-1 Minami-Osawa, Hachioji, Tokyo, Japan
         \and
         Department of Physics, Yamagata University, Kojirakawa-machi 1-4-12, Yamagata 990-8560, Japan
         \and
         Astronomical Institute Anton Pannekoek, University of Amsterdam, Science Park 904, 1098XH Amsterdam, the Netherlands
         \and
         GRAPPA Institute, University of Amsterdam, 1098 XH Amsterdam, The Netherlands
             }

   \date{}

  \abstract{
  We present the results of deep 140 ks {\it Suzaku} X-ray observations of the north-east (NE) radio relic of the merging galaxy cluster Abell\,2255. The temperature structure of Abell\,2255 is measured out to 0.9 times the virial radius (1.9 Mpc)  in the NE direction for the first time. The {\it Suzaku} temperature map of the central region suggests a complex  temperature distribution, which agrees with previous work.  
Additionally, on a larger-scale, we confirm that the temperature drops from 6 keV around the cluster center to 
3 keV at the outskirts, 
with two discontinuities at {\it r}$\sim$5\arcmin~(450 kpc) and $\sim$12\arcmin~(1100 kpc) from the cluster center. Their locations coincide with surface brightness discontinuities marginally detected in the XMM-Newton image, which indicates the presence of shock structures.
From the temperature drop, we estimate the Mach numbers  to be ${\cal M}_{\rm inner}\sim$1.2 and, ${\cal M}_{\rm outer}\sim$1.4. The first structure is most likely related to the large cluster core region ($\sim$350--430 kpc), and its Mach number is consistent with the {\it XMM-Newton} observation~\cite[${\cal M}\sim$1.24:][]{sakelliou06}.
Our detection of the second temperature jump, based on the {\it Suzaku} key project observation, shows the presence of a shock structure across the NE radio relic. This indicates a connection between the shock structure and the relativistic electrons that generate radio emission. Across the NE radio relic, however, we find a significantly lower temperature ratio ($T_1/T_2\sim1.44\pm0.16$ corresponds to~${\cal M}_{\rm X-ray}\sim1.4$) than the value expected  from radio wavelengths, based on the standard diffusive shock acceleration mechanism ($T_1/T_2>$ 3.2 or ${\cal M}_{\rm Radio}>$ 2.8).
This may suggest that under some conditions, in particular the NE relic of A2255 case, the simple diffusive shock acceleration mechanism is unlikely to be valid,  and therefore,  more a sophisticated mechanism is required.
}
\maketitle

%%%%%%%%%%%%%%%%%%%%%%%%%%%%%%%
%%%%%%%%%%%%%%%%%%%%%%%%%%%%%%%
%%%%%%%%%%%%%%%%%%%%%%%%%%%%%%%
\begin{figure*}
\begin{tabular}{cc}
\begin{minipage}{.5\hsize}
\begin{center}
\includegraphics[width=1.\hsize]{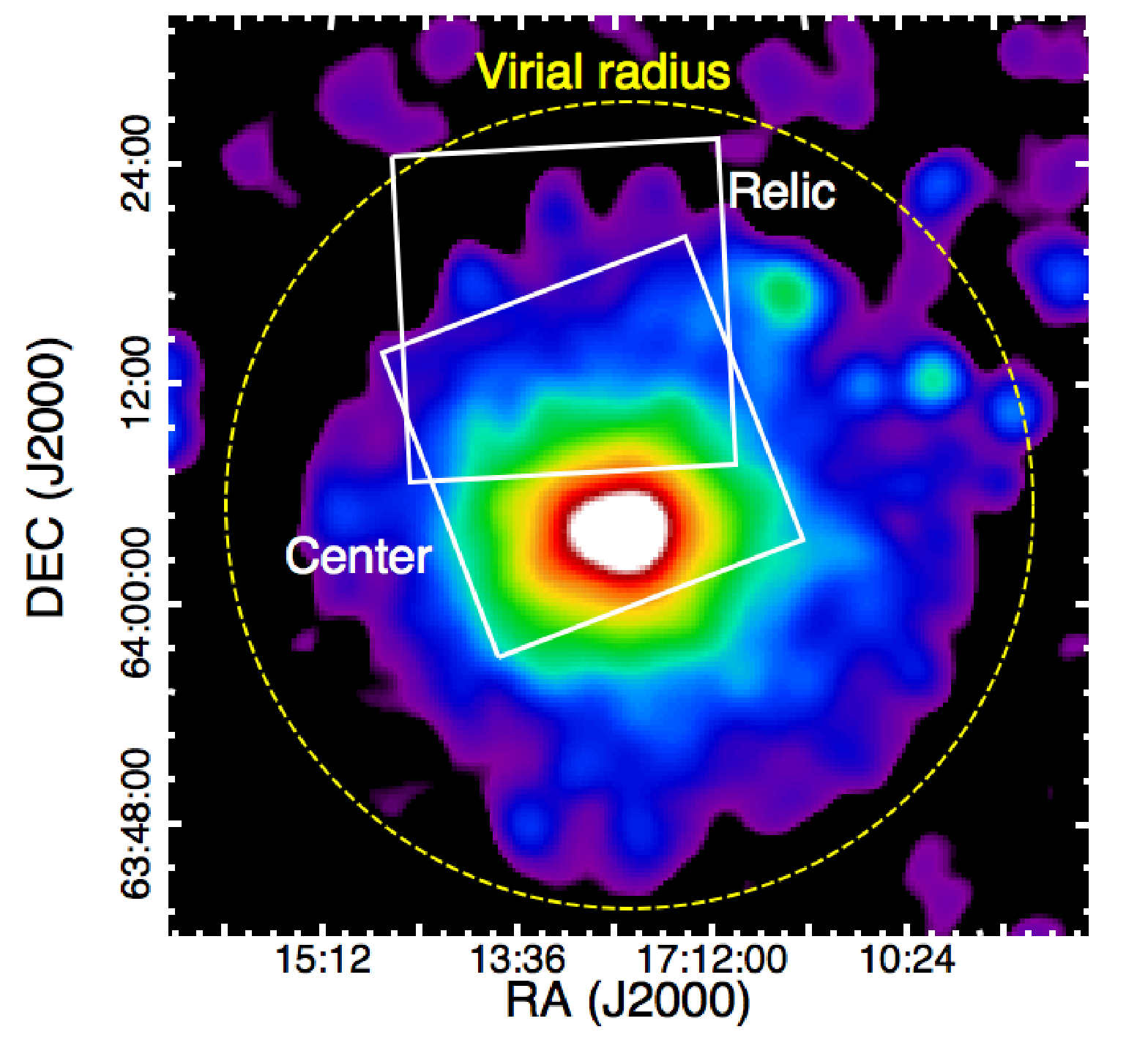}
\end{center}
\end{minipage}
\begin{minipage}{.5\hsize}
\begin{center}
\includegraphics[width=.8\hsize]{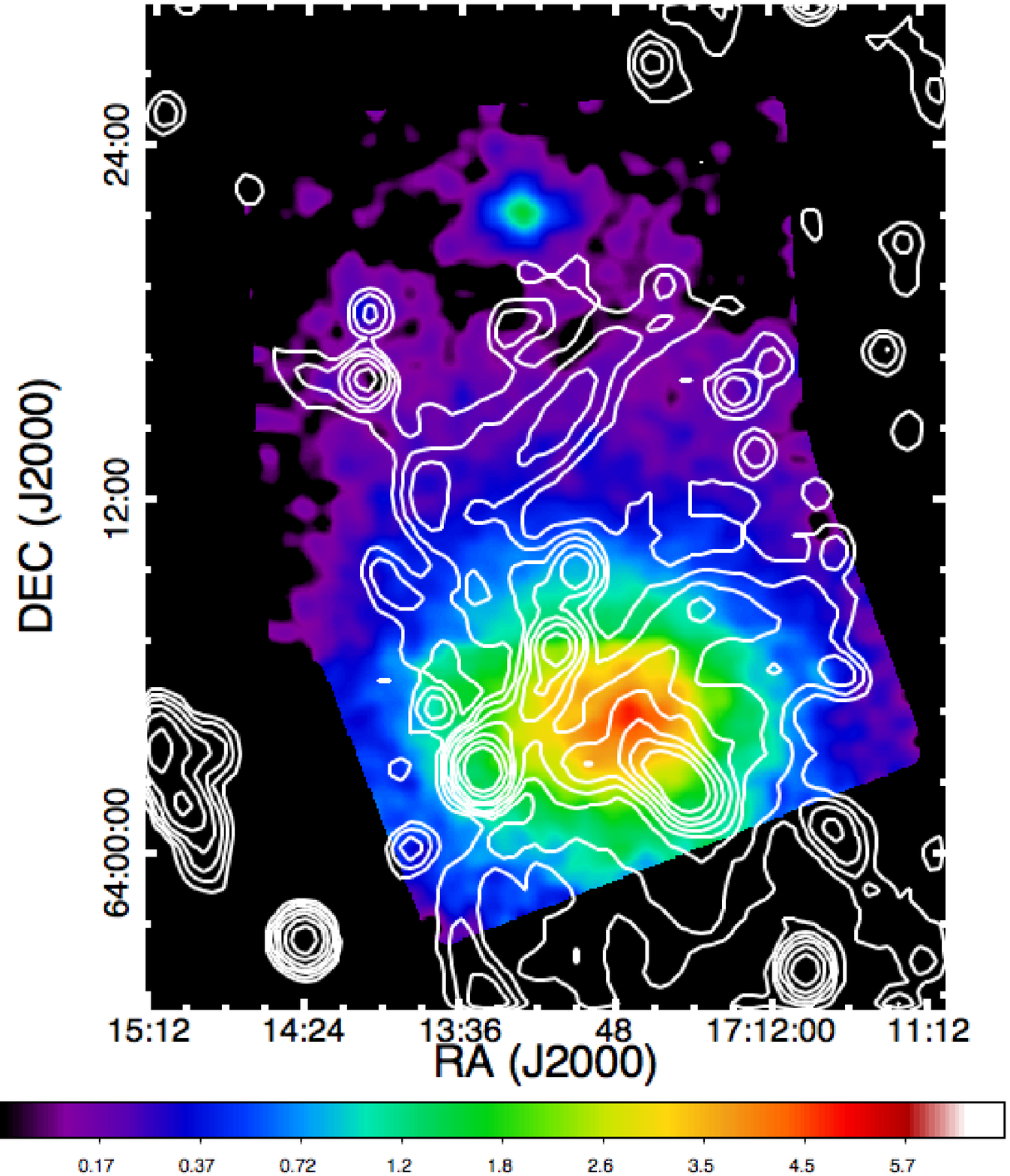}
\end{center}
\end{minipage}
\end{tabular}
\caption{\label{fig:suzaku_image}
Left: ROSAT image in the 0.1--2.4 keV band of A2255. The white boxes show the FOVs in the {\it Suzaku} XIS observations that we discuss in this paper, while the yellow dashed circle shows the virial radius of A2255, $r_{\rm 200}$, corresponding to 23.4\arcmin~ from the cluster center (see the text for details). 
Right: Background-subtracted {\it Suzaku} XIS3 image of A2255 in the 0.5--4.0 keV band smoothed by a two-dimensional Gaussian with $\sigma=8"$. The image was corrected for exposure time, but not for vignetting.  
The white contours are 1.0, 2.5, 5.0, 10, 20, 40, 80, and 160 mJy/beam of the WSRT 360 MHz radio intensity~\citep{pizzo09} .
}
\end{figure*}

%%%
%%% Introduction
%%%
\section{Introduction}\label{sec:intro}
Abell\,2255 (hereafter A2255) is a relatively nearby \cite[{\it z} = 0.0806:][]{struble99} rich merging  galaxy cluster.  Previous X-ray observations revealed that, the interacluster medium (ICM) has a complex temperature structure, suggesting that A2255 has experienced a recent, violent merger process~\citep{david93, burns95, burns98, feretti97, davis03,sakelliou06}. \citet{jones84} found that A2255 has a very large core region ({\it r$_c$}$\sim$350--430 kpc: see Fig.4 in their paper), which was confirmed by subsequent observations~\citep{feretti97,sakelliou06}. 

The most striking feature of A2255 is the diffuse complex radio emission~\citep{jeffe79, harris80, burns95,feretti97,govoni05, pizzo09}. 
This emission can be  roughly  classified into two categories: radio halos, and radio relics \cite[see Fig.5 in][]{ferrari08}. \citet{govoni05} discovered polarized filamentary radio emission in A2255.  \citet{pizzo08, pizzo09} revealed large-scale radio emission at the peripheral regions of the cluster and performed spectral index studies of the radio halo and relic. Information about the spectral index provides  important clues concerning the formation process of the radio emission regions. Basic details of the northeast radio relic (hereafter NE relic) are summerized in Table~\ref{tab:pizzo}. 
Several mechanisms have been proposed for the origin of the diffuse radio emission, such as turbulence acceleration and hadronic models for radio haloes, and diffusive shock (re-)acceleration \cite[DSA: e.g.,][]{bell87, blandford87} for radio relics~\cite[for details see~][and references therein]{brunetti14}.

Important open questions concerning radio relics include (1) why the Mach numbers of the shock waves inferred from X-ray and radio observation are sometimes inconsistent with each other~\citep[e.g.,][]{trasatti15, itahana15}, and (2) how weak shocks with ${\cal M}$ $<$ 3 can accelerate particles from thermal pools to relativistic energies via the DSA mechanism~\citep{kang12,pinzke13}. In some systems, the Mach numbers inferred from X-ray observations clearly violate predictions that are based on a pure DSA theory
\citep[][]{vink14}.

\begin{table}[t]
\caption{Expected shock properties associate with the NE radio relic from radio observation \citep{pizzo09}}
\centering
\begin{tabular}{ccccccccc}\hline 
 & Spectral &Mach number &  Expected {\it T} \\ 
 & index~$\alpha^\ast$& ${\cal M}_{\rm radio}^{\dagger}$ & ratio$^{\ddagger}$ \\
 \hline
{\rm 85~cm}-{\rm 2~m} & $0.5\pm0.1$ 	&  $>4.6$  & $>$ 7.5
\\
 {\rm 25~cm}--{\rm 85~cm}& $0.8\pm0.1$	& 2.77$\pm$0.35 & 3.2  
\\ \hline
\multicolumn{4}{l}{$\ast$: $S\propto\nu^{-\alpha}$, here we refer integrated spectral index} \\
\multicolumn{4}{l}{$\dagger$: Estimated from ${\cal M}_{\rm radio}=\frac{2\alpha+3}{2\alpha+1}$} \\
\multicolumn{4}{l}{$\ddagger$: Estimated from 
the Rankine-Hugoniot  jump condition
}\\
\end{tabular}
\label{tab:pizzo}
\end{table}
\begin{table*}
\caption{\label{tab:obslog}
Observation log and exposure time after data screening}
\begin{center}
\begin{tabular}{lllllll}  \hline   
Observatory &Name & Sequence ID& Position (J2000.0)& Observation  & Exp.$^{a}$ & Exp.$^{b}$ \\ 
& & & (R.A., DEC) & starting date & (ksec) & (ksec) \\
\hline  

{\it Suzaku}&Center & 804041010 & (258.24, 64.16) & 2010-02-07 & 44.5 & 41.6   \\
&Relic & 809121010 & (258.28, 64.26) & 2014-06-02 & 100.8 & 91.3$^{c}$  \\
&OFFSET &800020010 & (249.94, 65.20) & 2005-10-27 & 14.9 & 11.7  \\
\hline
{\it XMM-Newton} & XMM-C & 0112260801 & (258.19, 64.07)  &2002-12-07 & 21.0 & 9.0 \\
 & XMM-R &0744410501 &  (258.18, 64.42) & 2014-03-14 & 38.0  &  25.0
\\
\hline 
 \multicolumn{5}{l}{{\it a}: Without data screening} \\ 
 \multicolumn{5}{l}{{\it b}: {\it Suzaku}  data screening with COR2  $>$ 6 GV} \\ 
 \multicolumn{5}{l}{{\it c}: Additional processing for the XIS1 detector is applied (see text for details).} 
\end{tabular}
\end{center}
\end{table*}

Because X-ray observations enable us to probe ICM properties, a multiwavelength approach is a powerful tool to investigate the origin of the diffuse radio emission. However, because of the off-center location of radio relics in the peripheral regions of the cluster and the corresponding faint ICM emission, 
it has remained challenging to characterize them at  X-ray wavelengths \cite[e.g,][]{ogrean13}, except for a few exceptional cases under favorable conditions~\cite[e.g., Abell\,3667:][]{finoguenov10}.
The Japanese X-ray satellite {\it Suzaku}~\citep{mitsuda07}  improved this situation because of its low and stable background. {\it Suzaku} is a suitable observatory  to investigate low X-ray surface brightness regions such as cluster peripheries: see \citet[][]{hoshino10, simionescu11,kawaharada10,  akamatsu11} and \citet[][]{reiprich13} for a review.
To establish a clear picture of the detailed physical processes associated with radio relics, we conducted deep observations using {\it Suzaku} ({\it Suzaku} AO9 key project: Akamatsu et al.). In this paper, we report the results of {\it Suzaku} X-ray investigations on the NE relic in A2255.

We assume the cosmological parameters $H_0 = 70$ km s$^{-1}$ Mpc$^{-1}$, $\Omega_{\rm M}=0.27$ and $\Omega_\Lambda = 0.73$.
With a redshift of {\it z}=0.0806, 1\arcmin~corresponds to a diameter of 91.8 kpc. 
The virial radius,  which is represented by $r_{200}$, is approximately 
\begin{equation}
{\it r}_{200} = 2.77 h_{70}^{-1} (\langle T\rangle /10 \, {\rm keV})^{1/2}/E(z)\ {\rm Mpc}, 
\end{equation}
where {\it E(z)}=$(\Omega_{\rm M}(1+z)^{3}+1-\Omega_{\rm M})^{1/2}$~\citep{henry09}.  For our cosmology  and an average temperature of $\langle kT \rangle = 6.4$ keV (in this work: see Sect. \ref{sec:5arcmin})
{\it r}$_{200}$= 2.14 Mpc, corresponding to ${23.2\arcmin}$.
Here, we note that the estimated virial radius is generally consistent with the radius from HIFLUGCS \cite[$r_{200}=2.27^{+0.08}_{-0.07}$  Mpc:][]{reiprich02}.
For the comparison of the scaled temperature profiles (Sect. 4.1), we adopt the value given by {\it r}$_{200}$= 2.14 Mpc as the virial radius of A2255.
As our fiducial reference for the solar photospheric abundances  denoted by {\it Z}$_\odot$, we adopt the values reported by \citet{lodders03}. A Galactic absorption of $N_{\rm H}=2.7 \times 10^{20} \rm~cm^{-2}$~\citep{willingale13} was included in all fits.
Unless otherwise stated, the errors correspond to  68\% confidence for each parameter.

\begin{figure*}
\begin{tabular}{cc}
\begin{minipage}{.45\hsize}
\begin{center}
\includegraphics[width=1.1\hsize]{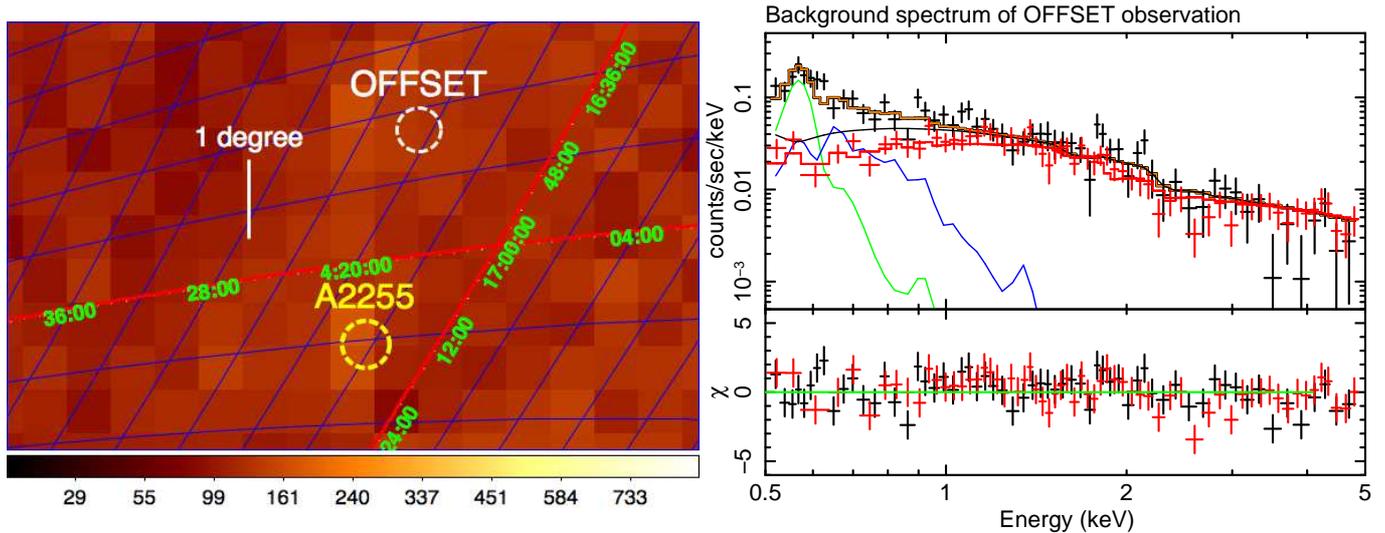}
\end{center}
\end{minipage}
\begin{minipage}{.55\hsize}
\begin{center}
\includegraphics[width=.7\hsize, angle=-90]{FIG4.ps}
\end{center}
\end{minipage}
\end{tabular}
\caption{\label{fig:bgd}
Left:  ROSAT R45 (0.47-1.20 keV) band image around A2255 in Galactic coordinates. The two circles ({\it r} = 20\arcmin) indicate the locations of the A2255 (yellow) and OFFSET (white dashed) observations.
Right: The NXB-subtracted XIS spectrum of the OFFSET observation.
The XIS BI (black) and FI (red) spectra are fitted with the sum of the CXB and the Galactic emission ({\it apec + phabs(apec + powerlaw)}). The CXB component is shown as a black curve, and the LHB and MWH  emissions are indicated with green and blue curves, respectively. 
}
\end{figure*}

\begin{table*}
\caption{\label{tab:bgd}Summary of the parameters of the fits for background estimation}
\begin{center}
\begin{tabular}{cccccccccccccc} \hline
 \multicolumn{2}{c}{LHB} &  \multicolumn{2}{c}{MWH} &  \multicolumn{2}{c}{CXB} & $\chi^2/d.o.f.$\\
 $kT$	& $norm^\ast$ ($\times10^{-2}$) &	 $kT$	& $norm^\ast$($\times10^{-4}$)  &	$\Gamma$	& $norm^\dagger$ \\ \hline
0.08 (fixed)	&  1.86$\pm0.36$	&0.25$\pm0.03$	&5.78$\pm0.19$& 1.41 (fixed)	& 10.3$\pm0.5$ & 142.9/118

\\ \hline
 \multicolumn{7}{l}{\footnotesize
$\ast$: Normalization of the apec component scaled 
by a factor 400$\pi$.}\\
\multicolumn{7}{l}{\footnotesize
$\dagger$: The
CXB intensity normalization in \citet{kushino02} is  
}\\
\multicolumn{7}{l}{
9.6$~\times10^{-4}$ for $\Gamma=1.41$
in units of photons keV$^{-1}~\rm cm^{-2}~s^{-1}$ at 1 keV.}
\end{tabular}
\end{center}
\end{table*}%

\section{Observations and data reduction}
{\it Suzaku} performed two observations: one aimed at the central region and  the other at the NE radio relic.  
Hereafter we refer to these pointings as Center and Relic, respectively (Fig.~\ref{fig:suzaku_image}). The X-ray imaging spectrometer \cite[XIS:][]{koyama07} on board {\it Suzaku} consists of three front-side illuminated (FI) CCD chips (XIS0, XIS2, and XIS3) and one back-side illuminated (BI) chip (XIS1).  After November 9, 2006, XIS2 was no longer operational  because of  damage from a micrometeoroid strike\footnote{http://www.astro.isas.jaxa.jp/suzaku/doc/suzakumemo/suzaku\\memo-2007-08.pdf}.  A similar accident occurred on the XIS0 detector\footnote{http://www.astro.isas.jaxa.jp/suzaku/doc/suzakumemo/suzakumemo-2010-01.pdf.}.
Because of flooding with a large amount of charges, segment A of XIS1 continuously saturated the analog electronics.  
For these reasons, we did not use data of XIS0. The segment A of XIS1 was also excluded.
All observations were performed with either the normal $5\times5$ or $3\times3$ clocking mode. Data reduction was performed with HEAsoft,  version 6.15, XSPEC version 12.8.1,  and CALDB version 20140624.

We started with the standard data-screening provided by the {\it Suzaku} team and applied event screening with a cosmic-ray cutoff rigidity (COR2) $>$ 6 GV to suppress the detector background. 
An additional screening for the XIS1 Relic observation data was applied to minimize the detector background. We followed the processes described in the {\it Suzaku} XIS official document\footnote{http://www.astro.isas.jaxa.jp/suzaku/analysis/xis/xis1\_ci\_6\_nxb/}.

We used {\it XMM-Newton} archival observations (ID:0112260801, 0744410501) for point source identification. We carried out the XMM-Newton EPIC data preparation following Sect. 2.2 in \citet{zhang09}. The resulting clean exposure times are 9.0 and 25.0 ks, respectively, and are shown in Table~\ref{tab:obslog}.

\section{Spectral analysis and results}\label{sec:spec}
\subsection{Spectral analysis approach}\label{sec:spec_model}
For the spectral analysis, we followed our previous approach as described in~\citet{akamatsu11, akamatsu12_a3667,akamatsu12_a3376}. 
In short, we used the following approach:
\begin{itemize}
\item Estimation of the sky background emission from the cosmic X-ray background (CXB), local hot bubble (LHB) and milky way halo (MWH) using a{\it Suzaku} OFFSET observations (Sect. \ref{sec:bgd}). 
\item Identification of the point sources using {\it XMM-Newton} observations and extraction from each {\it Suzaku} observation  (Sect. \ref{sec:bgd}).
\item With this background information, we investigate (i) the global properties of the central region of A2255 (Sect. \ref{sec:5arcmin}) and (ii) the radial temperature profile out to the virial radius (Sect. \ref{sec:radial}).
\end{itemize}

For the spectrum fitting procedure, we used the XSPEC package version 12.8.1. The spectra of the XIS BI and FI detectors were fitted simultaneously. For the spectral analysis, we generated the redistribution matrix file and ancillary response files assuming a uniform input image~({\it r} = 20\arcmin) by using {\it xisrmfgen} and {\it xisarfgen}~\citep{ishisaki07} in HEAsoft. The calibration sources were masked using the {\it calmask} calibration database file.

\subsection{Background estimation}\label{sec:bgd}
Radio relics are typically located in cluster peripheries ($\geq$~Mpc), where the X-ray emission from the ICM is faint. To investigate the ICM properties of such weak emission, an accurate and proper estimation of the background components is critical. The observed spectrum consists of the emission from the cluster, the celestial emission from non-cluster objects (hereafter sky background) and the detector background (non-X-ray background:\,NXB).

The sky background mainly comprises three components: LHB ($kT\sim$0.1 keV), MWH ({\it kT}$\sim$0.3 keV) and CXB.
To estimate them, we used a nearby {\it Suzaku} observation (ID: 800020010, exposure time of~$\sim$~12 ks after screening), 
which is located approximately 3\degr~from A2255. Figure~\ref{fig:bgd} (left) shows the ROSAT R45 (0.47-1.20 keV) band image around A2255. Most of the emission in this band is the sky background. 
Using the HEASARC X-ray Background Tool\footnote{http://heasarc.gsfc.nasa.gov/cgi-bin/Tools/xraybg/xraybg.pl}, the R45 intensities in the unit of 10$^6$ counts/sec/arcmin$^2$ are $115.5 \pm 2.4$ for a $r$=30\arcmin--60\arcmin ring centered on A2255 and $133.7 \pm 5.8$ for the offset region ($r$=30\arcmin circle), respectively. Thus they agree with each other within about 10\%.
We extracted spectra from the OFFSET observation and fitted them with a background model:
\begin{equation}
Apec_{\rm LHB}+phabs*(Apec_{\rm MWH}+Powerlaw_{\rm CXB}).
\end{equation}
The redshift and abundance of the two {\it Apec} components were fixed at zero and solar, respectively. We used spectra in the range of 0.5--5.0 keV for the BI and FI detectors. The model reproduces the observed spectra well with $\chi^2=142.9$ with 118 degrees of freedom. The resulting best-fit parameters are shown in Table~\ref{tab:bgd}, they are in good agreement with the previous study~\citep{takei07_a2218}.

For point-source identification in the {\it Suzaku} FOVs, we used {\it XMM-Newton} observations.  We generated a list of  bright point-sources using SAS task {\it edetect\_chain} applied to five energy bands 0.3--0.5 keV,  0.5--2.0 keV, 2.0--4.5 keV, 4.5--7.5 keV, and 7.5--12 keV using both EPIC pn and MOS data.
The point sources are shown as circles with a radius of $25^{\prime\prime}$ in Fig~\ref{fig:suzaku_image_reg};  this radius was used in the XMM-Newton analysis to estimate the flux of each source.

In the {\it Suzaku} analysis, we excluded the point sources with a radius of 1 arcmin  
to take the point-spread function (PSF) of the {\it Suzaku} XRT~\citep{xrt} into account. Additionally, we excluded a bright point-source to the north in the Relic observation  with a {\it r}=4\arcmin~ radius.

In Sect. \ref{sec:sys}, we investigate 
the impacts of systematic errors associated with the derived background model on the temperature measurement.

\begin{table}
\caption{Best-fit parameters for the central region ({\it r}$<$5\arcmin) of A2255 \label{tab:5arcmin}}
\begin{center}
\begin{tabular}{cccccccccccc} \hline
{\it kT} &  {\it Z} & Norm & $\chi^2$/d.o.f. \\
(keV)	    & ({\it Z}$_\odot$) & ($\times10^{-6}$) \\ \hline
$  {6.37}^{+0.06}_{-0.07}  $  & $ {0.28}^{+0.02}_{-0.02} $ & $ {167.1}^{+0.7}_{-0.9} $ & 1696 / 1657\\
\hline
\end{tabular}
\end{center}
\end{table}%
\begin{figure}
\begin{tabular}{c}
\begin{minipage}{\hsize}
\begin{center}
\includegraphics[width=.65\hsize,angle=-90]{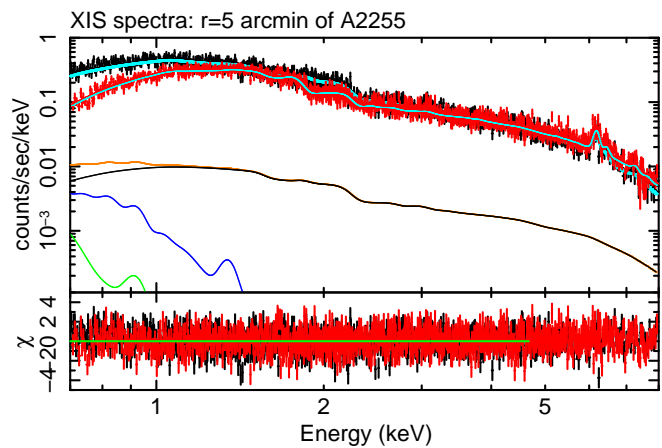}
\end{center}
\end{minipage}
\end{tabular}
\caption{\label{fig:5arcmin}
Left: NXB-subtracted spectra of A2255 ({\it r} $<$ 5\arcmin). The XIS BI (black) and FI (red) spectra are fitted with the ICM model ({\it phabs $\times$ Apec}), along with the sum of the CXB and Galactic emission. 
The ICM component is shown with the cyan line. The CXB, LHB and MWH emissions are indicated with black, green, and blue curves, respectively. The sum of all sky background emissions is shown with an orange curve. 
}
\end{figure}

\subsection{Global temperature and abundance}\label{sec:5arcmin}
First, we investigated the global properties (ICM temperature and abundance) of A2255. 
We extracted the spectra within {\it r} = 5\arcmin~of the cluster center~\cite[$\alpha=17^{h} 12^{m} 50^{s}.38, \delta=64d 03' 42''.6$:][]{sakelliou06} and fitted them with an absorbed thin thermal plasma model ({\it phabs~$\rm\times$~apec}) together with the sky background components discussed in the previous section (Sect. \ref{sec:bgd}). We kept the temperature and normalization of the ICM component as free parameters and fixed the redshift parameters to 0.0806~\citep{struble99}.
The background components were also fixed to their best-fit values
obtained in the OFFSET observation. 
For the fitting, we used the energy range of 0.7--8.0 keV for both detectors.
Figure~\ref{fig:5arcmin} shows the best-fit model.
We obtained fairly good fits with $\chi^2$=1696 with 1657 degrees of freedom (red $\chi^2=1.03$). The best-fit values are listed in Table~\ref{tab:5arcmin}. 

Substituting the global temperature of {\it} 6.37 keV into the $\sigma-T$ relation~(\cite{lubin93}:$~\sigma=10^{2.52\pm0.07}(kT)^{0.60\pm0.11}~\rm km~s^{-1}$), we obtain a velocity dispersion of $\sigma=1009^{+396} _{-302}~\rm km/s$, which is consistent with the values estimated by \cite{burns95}:1240$^{+203}_{-129}$ km/s,  \cite{yuan03}:1315$\pm$86~km/s and \cite{zhang11}: 998$\pm$55 km/s.

\begin{figure}
\begin{tabular}{c}
\begin{minipage}{1\hsize}
\includegraphics[width=\hsize]{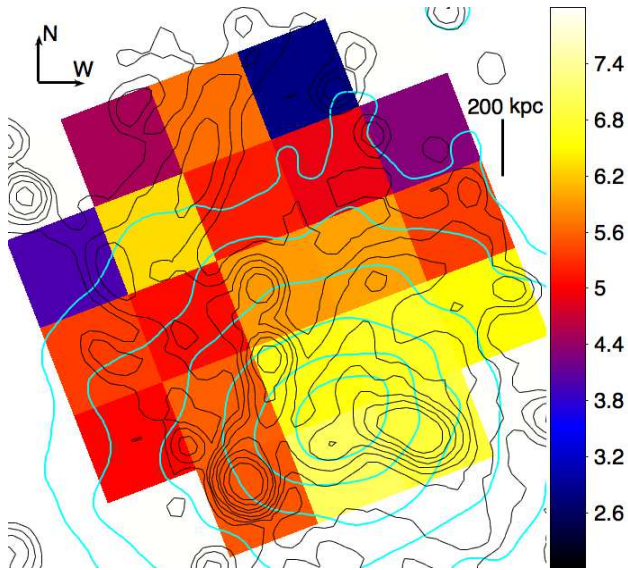}
\end{minipage}
\end{tabular}
\caption{\label{fig:map}
Temperature map of the central region of Abell 2255. The vertical color bar indicates the ICM temperature in units of keV. The lack of data in the four corners is due to the calibration source. The typical error for each box is about $\pm 0.6 $ keV. The cyan and black contours represent X-ray ({\it XMM-Newton}) and radio surface brightness distributions.
}
\end{figure}

To visualize the  temperature structure in the central region, we divided the Center observation into 5$\times$5 boxes and fitted them in the same manner as described above. 
Because A2255 is a merging cluster, we do not expect a strong abundance peak in the central region \citep{matsushita11}.  Therefore, the abundance for the central region was fixed to the global abundance value (0.3~{\it Z}$_\odot$) throughout the fitting. The resulting temperature map is shown in Fig.~\ref{fig:map}.
We note that the results are consistent
when the abundance is set  free.
The general observed trend matches previous work~\citep{sakelliou06}: the east region shows somewhat lower temperatures ({\it kT}$\sim$ 5 keV) than the west region ({\it kT}$\sim$ 7--8 keV).

The global temperature ({\it kT}=$6.37\pm0.07$ keV) and abundance ({\it Z}=0.28$\pm$0.02$Z_\odot$) agree with most previous studies~\cite[{\it EINSTEIN}, {\it ASCA}, {\it XMM-Newton}, and {\it Chandra}:][]{david93, ikebe02, sakelliou06, cavagnolo08}. However, {\it ROSAT} observations suggest a significantly lower temperature~\cite[$kT\sim$2--3 keV:][]{burns95, burns98, feretti97} than our results.  \citet{sakelliou06} discussed possible causes for such a disagreement, such as the difference of the source region, energy band, and the possible effect of a multi-temperature plasma.  On the other hand, investigations into this disagreement are beyond the scope of this paper.

\begin{figure}
\begin{tabular}{c}
\begin{minipage}{1\hsize}
\includegraphics[width=.6\hsize,angle=-90]{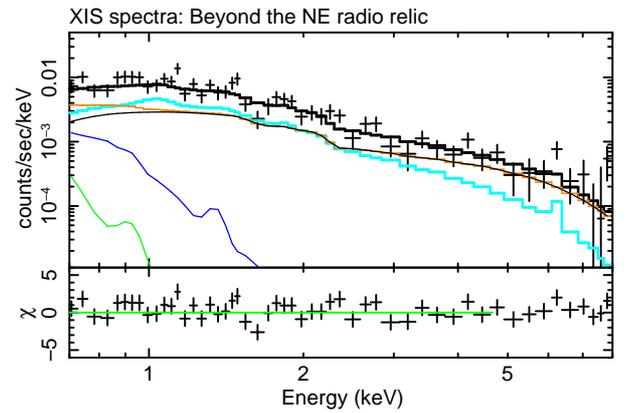}
\end{minipage}
\end{tabular}
\caption{\label{fig:spec_relic}
Same as Fig.~\ref{fig:5arcmin}, but beyond the NE radio relic.
For clarity, only the BI components are plotted.
}
\end{figure}

\begin{figure*}
\begin{tabular}{cc}
\begin{minipage}{.5\hsize}
\includegraphics[width=\hsize]{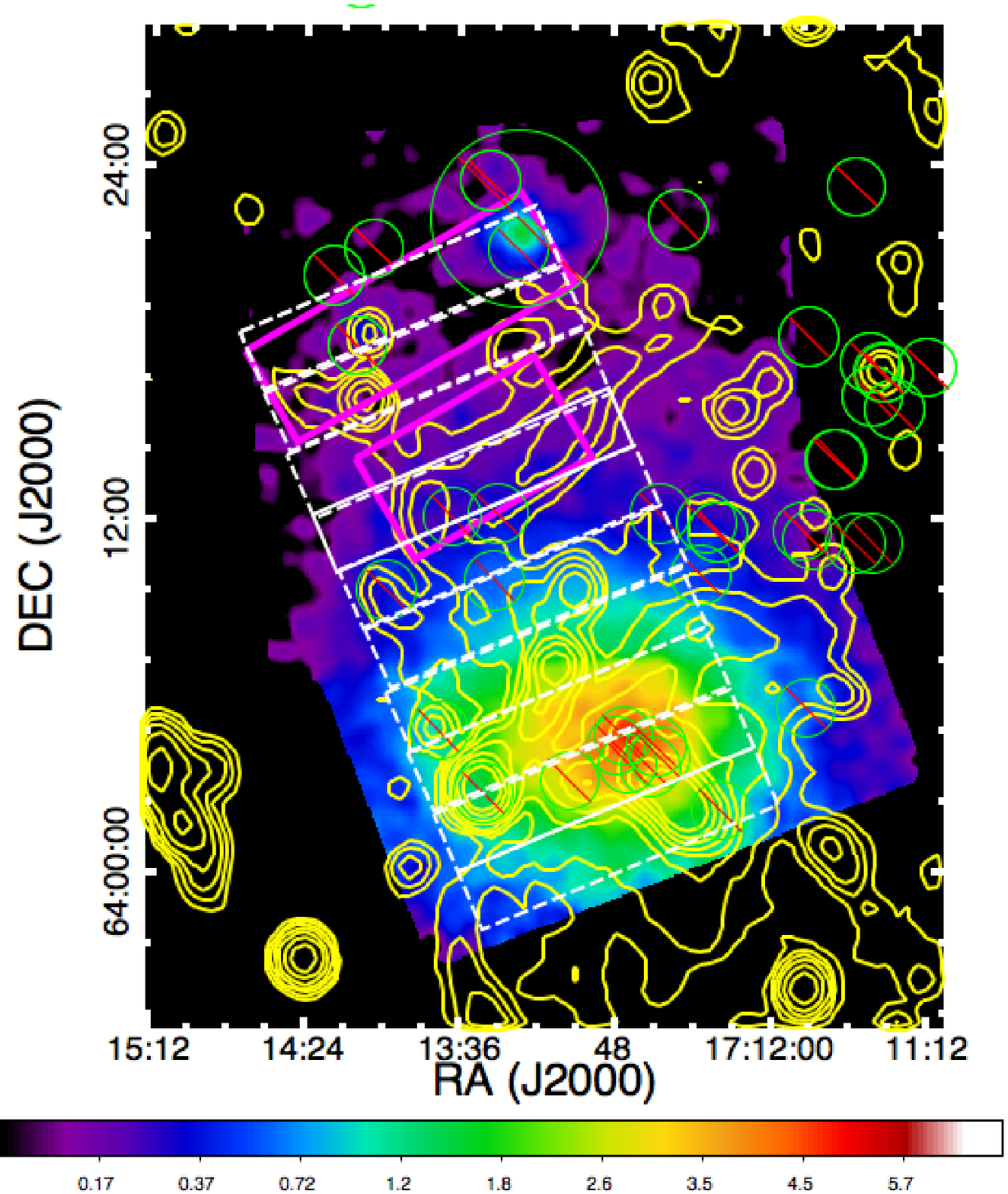}
\end{minipage}
\begin{minipage}{.5\hsize}
\includegraphics[width=\hsize]{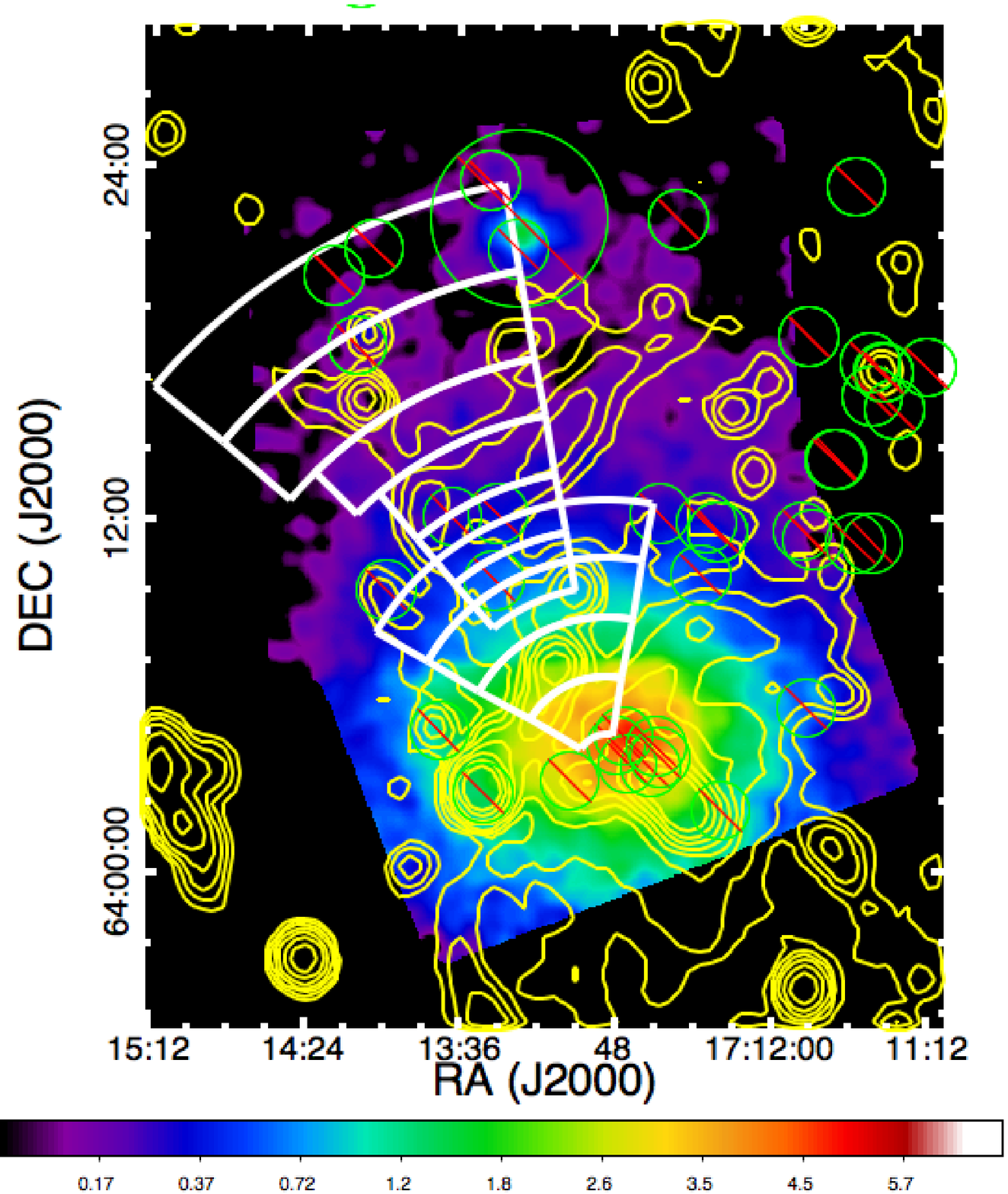}
\end{minipage}
\end{tabular}
\caption{\label{fig:suzaku_image_reg}
XIS image of A2255. The spectral regions used in the spectral analysis are indicated with the white boxes (2\arcmin$\times$10\arcmin) (left) and sectors (right).  In the left panel, the magenta boxes show the regions used for the pre- and post-shock region.
The point sources, identified by XMM-Newton, are highlighted with the green circles (see text for details).
}
\end{figure*}

\begin{figure*}
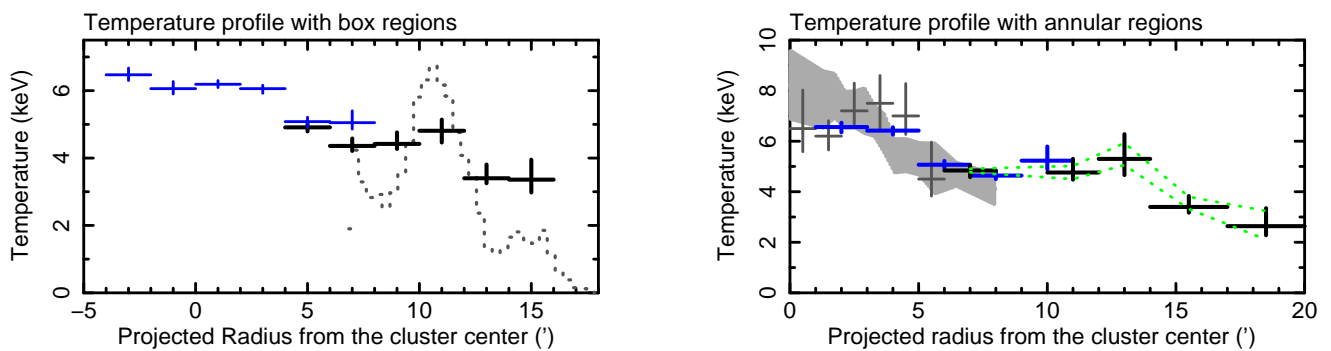

\begin{tabular}{cc}
\begin{minipage}{.5\hsize}
\includegraphics[width=.5\hsize,angle=-90]{FIG10.ps}
\end{minipage}
\begin{minipage}{.5\hsize}
\includegraphics[width=.5\hsize,angle=-90]{FIG11.ps}
\end{minipage}
\end{tabular}
\caption{\label{fig:box}
Left: The temperature profile of A2255 in the direction from south to NE. Blue and black crosses indicate the best-fit temperature based on the Center and Relic observations, respectively. The dotted gray lines show the profile of the radio emission taken from the WSRT radio data~\citep{pizzo09}. The discrepancy between the Center and Relic observations at {\it r}=7\arcmin is due to the lack of the XIS0 data of the Center observation.
Right:
Radial temperature profile of A2255 in the  NE radio relic direction. The gray crosses and the dark gray region indicate the radial profiles obtained by {\it XMM-Newton}~\citep{sakelliou06} and {\it Chandra}~\citep{cavagnolo09}, respectively. Green dashed lines show typical systematic changes in the best-fit values that are due to changes in the CXB intensity and NXB level. 
}
\end{figure*}

\subsection{Radial temperature profile for the NE direction}\label{sec:radial}
\begin{table}
\caption{Flux ratio of the ICM and the sky background components beyond the relic and outermost bins\label{tab:ratio}}
\begin{center}
\begin{tabular}{ccc|cccccc} \hline
			&\multicolumn{2}{c|}{Box}& Sector \\ \hline
			& Beyond the relic$^{\ast}$	&	 Outermost 		& Outermost\\ 
	&  (0.60--0.71)$r_{200}$		&	(0.71--0.81)$r_{200}$	& (0.75--0.85)$r_{200}$\\		\hline
0.7--2.0 keV	&	1.09				&		1.15				&	0.92	\\
0.7--8.0 keV	&	0.79				&		0.67				&	0.48	\\ 
\hline
\multicolumn{4}{l}{ 
$\ast$: Big magenta box in the left panel of Fig.~\ref{fig:suzaku_image_reg}.
}
\end{tabular}
\end{center}
\end{table}%

\begin{table}
\caption{\label{tab:radial} Best-fit parameters for the radial regions.}
\begin{center}
\begin{tabular}{cccccccccccc} \hline
&Radius (\arcmin)	& {\it kT} (keV)	& Norm	& $\chi^2/d.o.f.$ \\ \hline
&2.0 $ \pm $ 1.0 & $  {6.56}^{+0.17}_{-0.21}  $   & $ {196.5}^{+3.1}_{-3.8}$  & 201 / 205\\
&4.0 $ \pm $ 1.0 & $  {6.42}^{+0.13}_{-0.16}  $   & $ {122.6}^{+1.3}_{-1.6}$  & 289 / 297\\
Center&6.0 $ \pm $ 1.0 & $  {5.07}^{+0.15}_{-0.15}  $   & $ {66.0}^{+0.7}_{-0.8}$  & 279 / 243\\
&8.0 $ \pm $ 1.0 & $  {4.64}^{+0.23}_{-0.22}  $   & $ {34.9}^{+0.5}_{-0.9}$  & 101 / 138\\
&10.0 $ \pm $ 1.0 & $  {5.23}^{+0.50}_{-0.34}  $   & $ {18.4}^{+0.5}_{-0.7}$  & 109 / 82\\ 
\hline
&7.0 $ \pm $ 1.0 & $  {4.84}^{+0.20}_{-0.26}  $   & $ {44.3}^{+0.7}_{-1.1}$  & 129 / 99\\
&9.0 $ \pm $ 1.0 & $  {4.13}^{+0.21}_{-0.23}  $   & $ {27.2}^{+0.7}_{-0.8}$  & 89 / 67\\
Relic&11.0 $ \pm $ 1.0 & $  {4.76}^{+0.38}_{-0.30}  $   & $ {11.5}^{+0.4}_{-0.4}$  & 125 / 93\\
&13.0 $ \pm $ 1.0 & $  {4.89}^{+0.30}_{-0.35}  $   & $ {8.0}^{+0.3}_{-0.4}$  & 96 / 81\\
&16.5 $ \pm $ 2.0 & $  {3.42}^{+0.48}_{-0.15}  $   & $ {5.1}^{+0.2}_{-0.3}$  & 131 / 124\\
\hline
\end{tabular}
\end{center}
\end{table}%

\begin{table}
\caption{\label{tab:relic} Best-fit parameters for the pre- and post-shock regions}
\begin{center}
\begin{tabular}{cccccccccccc} \hline
	& {\it kT} (keV)	& Norm	& $\chi^2/d.o.f.$ \\ \hline
Post	&  $  {4.52}\pm0.36  $   & $ {7.6}\pm0.2$  & 198 / 157 \\
Pre	&  $  {3.34}\pm0.46  $   & $ {4.8}\pm0.3$  & 114 / 91
\\
\hline
\end{tabular}
\end{center}
\end{table}%

To investigate the temperature structure of A2255 toward the NE relic with its radio emission, we extracted 10 boxes (2\arcmin$\times$10\arcmin), as shown in Fig.\ref{fig:suzaku_image_reg} (left). We followed the same approach as described above, and fixed the abundance to 0.3~{\it Z}$_\odot$, 
which is consistent with the abundance of the central region (previous section) and the measured outskirts of other clusters~\citep{fujita08, werner13}. 
We successfully detect the ICM emission beyond the NE radio relic.  Table~\ref{tab:ratio} shows the ratio of the ICM and the sky background components at the outer regions. 
These values are consistent with other clusters~\citep[e.g. one could derive the ratio of the ICM and the X-ray background from the right bottom panel of Fig.~2 in][]{miller12}.
In general, we obtained good fits for all boxes ($\chi^{2}/d.o.f. <1.2$). The resulting temperature profile is shown in Fig.~\ref{fig:box}, where the blue and black crosses indicate the best-fit temperature derived from the Center and Relic observations, respectively. The gray dotted histogram shows the  profile of the radio emission taken from the WSRT radio data~\citep{pizzo09}. The temperature profile changes from {\it kT}$\sim$ 6 keV at the center to $\sim$3 keV beyond the NE radio relic. Furthermore, there are two distinct temperature drops at {\it r} = 4\arcmin~ and 12\arcmin, respectively. The latter structure seems to be correlated with the radio relic, which is consistent with other systems~\cite[e.g., A3667:][]{finoguenov10, akamatsu12_a3667}. In order to estimate the properties of the ICM across the relic, we extracted spectra from the regions indicated by magenta boxes in Fig.~\ref{fig:suzaku_image_reg} (left). We selected the beyond the relic region with 1\arcmin separation to avoid a possible contamination from the bright region. The best-fit values are summarized in Table~\ref{tab:relic}, which are well consistent with the results of box-shaped region. We evaluate the shock properties related to the NE radio relic in the discussion (Sect. \ref{sec:shock} and \ref{sec:comp}).

To investigate the former structure and radial profile, we extracted spectra by selecting annuli from the center to the periphery as shown in Fig.~\ref{fig:suzaku_image_reg} (right). The radial temperature profile also shows clear discontinuities at {\it r} = 5\arcmin~and 14\arcmin~as shown Fig.~\ref{fig:box}. In Fig.~\ref{fig:box}, we compare the {\it Suzaku} temperature profile with {\it Chandra}~\citep{cavagnolo09} and {\it XMM-Newton} profiles~\citep{sakelliou06}. Both results covered the central region, where they agree well with the {\it Suzaku} result.

\subsection{Systematic uncertainty in the temperature measurement}\label{sec:sys}
Because the X-ray emission around radio relics is relatively weak, it is important to evaluate the systematic uncertainties in the temperature measurement. First we consider the systematic error of estimating the CXB component in the Relic observation and the detector background.
In order to evaluate the uncertainty of the CXB that is due to the statistical fluctuation of the number of point sources, we followed the procedure described in \citet{hoshino10}.  
In this paper we refer to the flux limit
 as $1\times10^{-14} \rm~erg/s/cm^2$ determined by XMM-Newton. The resulting CXB fluctuations span 11--27\%. 
We investigate the impact of this systematic effect by changing the intensity by $\pm$11--27 \% and the normalization of the NXB by $\pm$ 3\%~\citep{tawa08}.

The resulting best-fit values, after taking into account the systematic errors, are shown with green dashed lines in Fig.~\ref{fig:box}. The effects of the systematic error that is due to the CXB and NXB are smaller than or compatible to the statistical errors. This result is expected because the regions  discussed here are  within  the virial radius ($<0.8r_{200}$) and 
 the ICM emission is still comparable to that of the sky background components even around the NE relic (Table~\ref{tab:ratio}).

Secondl, we investigated the effects of flickering pixels. Following the official procedure\footnote{http://www.astro.isas.ac.jp/suzaku/analysis/xis/nxb\_new/}, we confirmed that the effect of the flickering pixels was smaller than the statistical error. 

Third, because the spatial resolution of the {\it Suzaku} telescope is about 2\arcmin, the temperature profile is likely to be affected by the choice of region. To study this effect, we  shifted the boxes in Fig~\ref{fig:box} (left) by 1.5\arcmin~ (half of its size) outside and measured the temperature profile again: we found that the temperature ratio across the relic decreased by 20\%. 

\begin{figure}
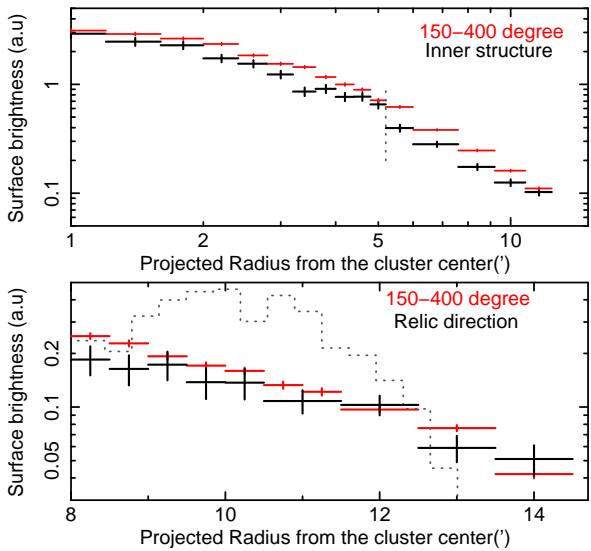

\begin{tabular}{c}
\begin{minipage}{1\hsize}
\includegraphics[width=.4\hsize, angle=-90]{FIG12.ps} \\
\includegraphics[width=.4\hsize, angle=-90]{FIG13.ps}
\end{minipage}
\end{tabular}
\caption{\label{fig:US}
{\it XMM-Newton} 0.5-1.4 keV surface brightness profiles (top: inner temperature structure, bottom: across the radio relic). Red and black crosses represent the radial profile of an opening angle with 150\degr--400\degr and 80\degr--130\degr, respectively. In the top plot, the vertical dashed line indicates the location of the discontinuity.  In the bottom plot, the dotted gray lines show WSRT 360 MHz radio emission.
} 
\end{figure}

\begin{figure}
\begin{tabular}{c}
\begin{minipage}{1\hsize}
\includegraphics[width=\hsize]{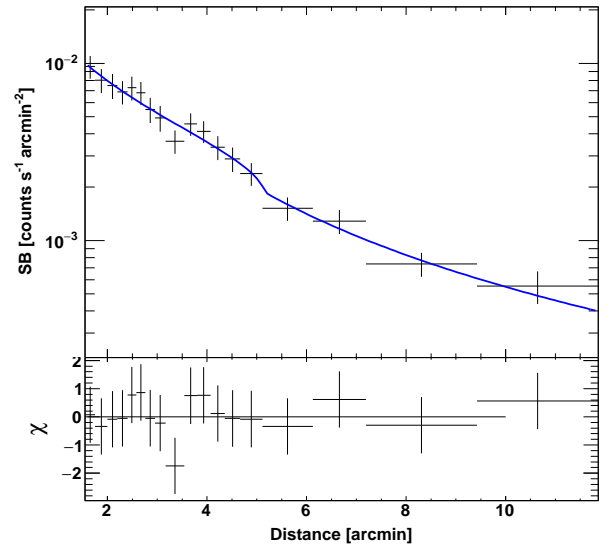}
\end{minipage}
\end{tabular}
\caption{\label{fig:proffit}
{\it XMM-Newton} 2.0-7.0 keV surface brightness profile.
The best-fit for the profile is plotted and their residuals are shown in the bottom panel.
The inner jump was marginally detected at {\it r}=5.2\arcmin$\pm$0.6\arcmin.
}
\end{figure}

\subsection{Surface brightness structures}\label{sec:sb}
In order to investigate whether the observed temperature structures
are shocks or cold fronts,
 we extracted the surface brightness profile from the 0.5--1.4 keV {\it XMM-Newton} image 
 in the northeast sector  with an opening angle of 80\degr--130\degr.  The resulting surface  brightness profile is shown in the top (the inner structure)
and bottom (across the radio relic) panels of Fig. 8. 
In both panels, 
black and red crosses represent the radial profiles in the  80\degr--130\degr (NE) and 150\degr--400\degr sectors, respectively, where we use the latter as the control sector to be compared with the NE sector.

In the top panel of Fig. 8, the discontinuity in {\it XMM-Newton} surface brightness profile is clearly visible. 
The location of the discontinuity around {\it r}$\sim$5\arcmin.2, indicated by the dotted gray line, is consistent with that of the inner temperature structure.
We found a signature of the surface brightness drop across the relic ({\it r}$\sim$12\arcmin.5). 
By taking the ratio of the surface brightness across the discontinuity/drop, 
we estimated the ratio of the electron density $C\equiv n_2/n_1$ as $C_{\rm Inner}=1.28\pm0.07$ and $C_{\rm Outer}=1.32\pm0.14$, respectively.
We also found a marginal sign of the inner structure by using the {\tt PROFFIT} software package~\citep{proffit}. We assume that the gas density follows two power-law profiles connecting at a discontinuity with a density jump. In order to derive the best-fit model, the density profile was projected onto the line of sight with the assumption of spherical symmetry. We extracted in the same matter as described above. We obtained the best-fit ratio of the electron density as $C_{{\rm Inner}}$=1.33$\pm$0.24 with reduced $\chi^{2}$=0.53 for 12 d.o.f., which is in good agreement with the above simple estimation. The surface brightness profile and resulting best-fit model are shown in Fig.~\ref{fig:proffit}. 

Although the significance is not high enough to claim a firm detection, we consider it is reasonable to think that the inner structure is a shock front (at least not a cold front). For the relic, we were not able to detect a structure across the relic because of poor statistics. Cold fronts at cluster periheries are rare~\citep[see Fig.5 in ][]{walker14}, and the observed ICM properties across the relic are supportive of the presence of a shock front. 
Therefore, we conclude that the two temperature structures are shock fronts.
We discuss their properties in the following section.  

\subsection{Constraints on nonthermal emissions}\label{sec:IC}
To evaluate the flux of possible nonthermal X-ray emission caused by inverse Compton scattering by the relativistic electrons from the NE relic, we reanalyzed  the relic region by adding a power-law component to the ICM model described in the previous subsection.
Based on the measured radio spectral index~\citep{pizzo09}, we adopted $\Gamma_{\rm relic}$=$\alpha_{\rm relic}$ +1 = 1.8 (Table~\ref{tab:pizzo}) as a photon index for the NE relic. 
As mentioned above, the observed spectra are well modeled with a thermal component. Therefore, we could set an upper limit to the nonthermal emission. 
The derived one sigma (68\%) upper limit of the surface brightness in the 0.3--10 keV is  $F_{\rm relic}<1.8\times10^{-14}\rm~ erg~cm^{-2}~s^{-1}~arcmin^{-2}$.
Assuming the solid angle covered by the NE relic~\cite[$\Omega_{\rm NE~ relic}=60~\rm arcmin^2$, ][]{pizzo09}, 
the upper limit of the flux in the 0.3--10 keV band is $<1.0\times10^{-12}\rm~ erg~cm^{-2}~s^{-1}$. 
This upper limit is somewhat more stringent than that for the halo~\cite[see Fig.10 in][for  a review]{ota12}.  The reason for a more stringent upper limit is that for the periphery the lower surface brightness of thermal emission would cause the nonthermal emission to bemore prominent.

\section{Discussion}\label{sec:discussion}
The {\it Suzaku} observations of the merging galaxy cluster A2255 were conducted out to the virial radius ($r_{200}=2.14$ Mpc or 23.2\arcmin).  With its deep exposure, it allowed us to investigate the ICM properties beyond the central region. We successfully characterized the ICM temperature profile out to 0.9 times the virial radius. We also confirmed the temperature discontinuities in the NE direction, which indicate shock structures. In the following subsections, we discuss the temperature structure by comparing it to other clusters, evaluate the shock properties, and discuss their implications.

\begin{table*}
\caption{\label{tab:mach}Shock properties at the inner and outer structures}
\begin{center}
\begin{tabular}{cccccccc} \hline
 &$T_{1}$ &$T_{2}$ & $T_{1}/T_{2}$  &${\cal M}$  & Compression &  Propagation velocity\\
  & (keV) & (keV)  & & & & (km~s$^{-1}$)\\ \hline
  Inner  & 6.42$\pm0.14$ & 5.07$\pm0.15$ & 1.27$\pm0.04$ & 1.27$\pm$0.04 & 1.40$\pm0.05$ &1440$\pm$50
  \\
  Outer & 4.52$\pm$0.36 &3.34$\pm0.46$ & 1.35$\pm$0.16&  1.36$\pm0.16$ & 1.53$\pm$0.23 & 1380$\pm$200
  \\ \hline
\end{tabular}
\end{center}
\end{table*}%

\subsection{Shock structures and their properties}\label{sec:shock}

Here, we evaluated  the Mach number of each shock using the Rankine-Hugoniot jump condition~\citep{landau59_fluid}
\begin{equation}
\frac{T_2}{T_1} = \frac{5{\cal M}^4+14{\cal M}^2-3}{16{\cal M}^2},
\end{equation}
where 1 and 2 denote the pre- and post-shock regions, respectively. 
We assumed the ratio of specific heats to be $\gamma=5/3$. The estimated Mach numbers are shown in Table~\ref{tab:mach}, with values of ${\cal M}_{\rm inner}=1.27\pm0.06$ and ${\cal M}_{\rm outer}=1.36\pm0.16$.
Based on the Mach numbers, we estimated the shock compression parameter to be $C_{\rm Inner}=1.40\pm0.05$ and $C_{\rm Outer}=1.53\pm0.23$, respectively.
These values are consistent with those derived from the surface brightness distributions (Sect.~\ref{sec:sb}).
Using the measured pre-shock temperature, the sound speeds are $c_{\rm inner}\sim1170\rm~km~s^{-1}$ and $c_{\rm outer}\sim 950\rm~km~s^{-1}$. The shock propagation velocity can be estimated from
 $v_{\rm shock}=c_s\times{\cal M}$ with values of $v_{\rm shock:inner}=1440\pm50\rm~km~s^{-1}$ and $v_{\rm shock:outer}=1380 \pm 200\rm~km~s^{-1}$.
These shock properties are  similar to those observed in other galaxy clusters: Mach number ${\cal M}\sim1.5-3.0$ and shock propagation velocity $v_{\rm shock}\sim1200-3500~\rm km~s^{-1}$~\citep{markevitch02,markevitch05, finoguenov10,macario11,mazzotta11, Russell12, dasadia16}.

\begin{figure}
\begin{tabular}{c}
\begin{minipage}{1\hsize}
\includegraphics[width=1\hsize]{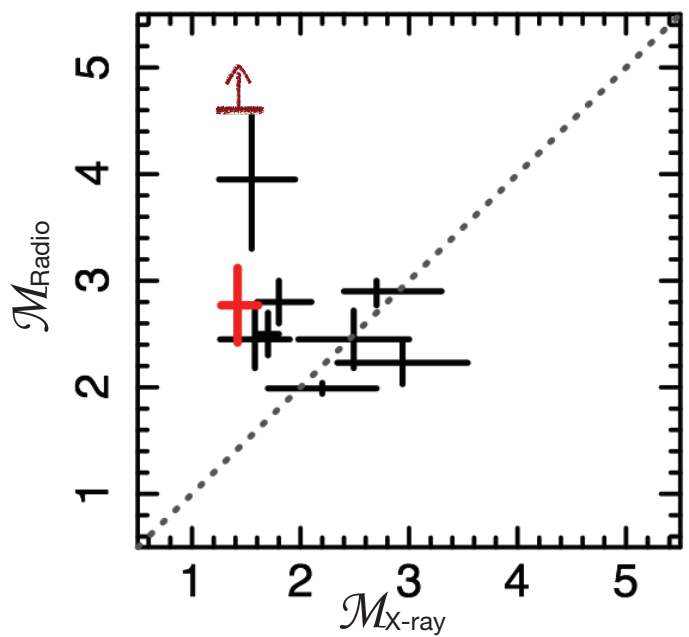}
\end{minipage}
\end{tabular}
\caption{\label{fig:mach_comp}
Mach number derived from radio observations (${\cal M}_{radio}$) plotted against that from the ICM temperature (${\cal M}_X$). The results of A2255 are indicated by 
the red cross (assuming radio spectral index: $\alpha_{\rm 85 cm}^{\rm 25 cm}=-0.8\pm0.1$) and the brown lower limit ($\alpha_{\rm 85 cm}^{\rm 2 m}=-0.5\pm0.1$). 
The black crosses show the results of other systems (see text for the detail).
The gray dashed line indicates the linear correlation as a reference.
Here we note that not all Mach numbers inferred from radio observations based on injected spectral index. Therefore, future low-frequency radio observations can change the Mach numbers displayed here (see text for a detail).
}
\end{figure}

\subsection{X-ray and radio comparison}\label{sec:comp}
Based on the assumption of simple DSA theory, the Mach number can be also estimated from the radio spectral index via
$\displaystyle{{\cal M}_{\rm radio}=\frac{2\alpha+3}{2\alpha+1}}$
\citep{dury83,blandford87}. 
From the observed radio spectral indices~\citep{pizzo09}, the expected Mach numbers at the NE relic are ${\cal M}_{\rm 85~cm-2~m}> 4.6$, ${\cal M}_{\rm 25~cm-85~cm}=2.77\pm0.35$, which are higher than our X-ray result (${\cal M}_{\rm X}=1.42^{+0.19}_{-0.15}$). 
Even when the systematic errors estimated in Sect.~\ref{sec:sys} are included, the difference between the Mach numbers inferred from the X-ray and radio observations still holds. 

One possible cause of this disagreement is that the {\it Suzaku} XRT misses or dilutes the shock-heated region because of its limited spatial resolution~\cite[HPD$\sim$1.7\arcmin:][]{xrt}. This would indicate that the {\it Suzaku} XIS spectrum at the post-shock is a multi-temperature plasma consisting of pre- and post-shock media~\citep{sarazin14}. To confirm this effect, we reanalyzed the post-shock region with a two-temperature ({\it 2 kT}) model by adding an additional thermal component to the single-temperature model of the post-shock region.  
However, there is no indication that another thermal component is needed for the post-shock spectrum. 
The single-temperature model reproduces the measured spectra well~(Fig.~\ref{fig:spec_relic}).
We note here that it is hard to evaluate the hypothesis of the two-phase plasma with the available X-ray spectra. And as pointed out by~\citet{kaastra04}, the current X-ray spectroscopic capability has its limit on the X-ray diagnostics of the multi-phase plasma. 
\citet{mazzotta04} 
also demonstrated the difficulties in distinguishing a two-temperature plasma when both temperatures are above 2 keV (see Figs.~1 and 3 in their paper).

Relativistic electrons just behind the shock immediately (within $\sim10^7$ years) lose their energy via radiative cooling and energy loss via inverse-Compton scattering. Since the cooling timescale of electrons is shorter than the lifetime of the shock, spectral curvature develops, which is called the ``aging effect''. As a result, the index of the integrated spectrum decreases at the high-energy end by about 0.5 from $\alpha_{\rm inj}$ for a simple DSA model~\cite[$\alpha_{\rm int}=\alpha_{\rm inj}-0.5$:~e.g., ][]{pacholczyk70, miniati02}. 
This means that high-quality and spatially resolved low-frequency observations are  needed to obtain $\alpha_{inj}$ directly.
 The number of relics with measured $\alpha_{inj}$ with  enough sensitivity and angular resolution 
 is small because of the challenges posed by low-frequency radio observations~\cite[e.g.,][]{vanweeren10, vanweeren12_toothbrush, vanweeren16_toothbrush}.
For the NE relic in A2255,  \citet{pizzo09} confirmed a change in the slope with a different frequency band $(\alpha_{\rm 25~cm}^{\rm 85~cm}=-0.8\pm0.1$ and $\alpha_{\rm 85~cm}^{\rm 2~m}=-0.5\pm0.1$), indicating the presence of the aging effect. However, their angular resolution is not good enough to resolve the actual spectral index before it affects the aging effect.
Therefore, the actual Mach number from the radio observations might be different from 
what is discussed here.

Although the X-ray photon mixing and radio aging can have some effect, it is still worthwhile to compare the A2255 results with those of other clusters with radio relic. Figure~\ref{fig:mach_comp} shows a comparison of the Mach number inferred from X-ray and radio observations, which were taken and updated from~\citet{akamatsu13a}. The result for the NE  relic of A2255 is shown by the red cross together with other objects: CIZA J2242.8+5301 south and north relics~\citep{stroe14c, akamatsu15}, 1RXS J0603.3+4214~\citep{vanweeren12_toothbrush, itahana15}, A3667 south and north relics~\citep{hindson14, akamatsu12_a3667, akamatsu13a}, A3376~\citep{kale12, akamatsu12_a3376}, the Coma relic~\citep{thierbach03, akamatsu13b} and A2256~\citep{trasatti15}. 
This discrepancy, ${\cal M}_{\rm radio}> {\cal M}_{\rm X}$, has also been observed for other relics (A2256, Toothbrush, A3667 SE).  If this discrepancy is indeed real, it may point to
problems in the basic DSA scenario for shocks in clusters.  In other words, at least for these objects, the simple DSA assumption does not hold, and therefore, a more complex mechanism might be required. We discuss possible scenarios for this discrepancy in the next subsection.

\begin{figure*}
\begin{tabular}{cc}
\begin{minipage}{.5\hsize}
%\begin{center}
\includegraphics[width=1\hsize]{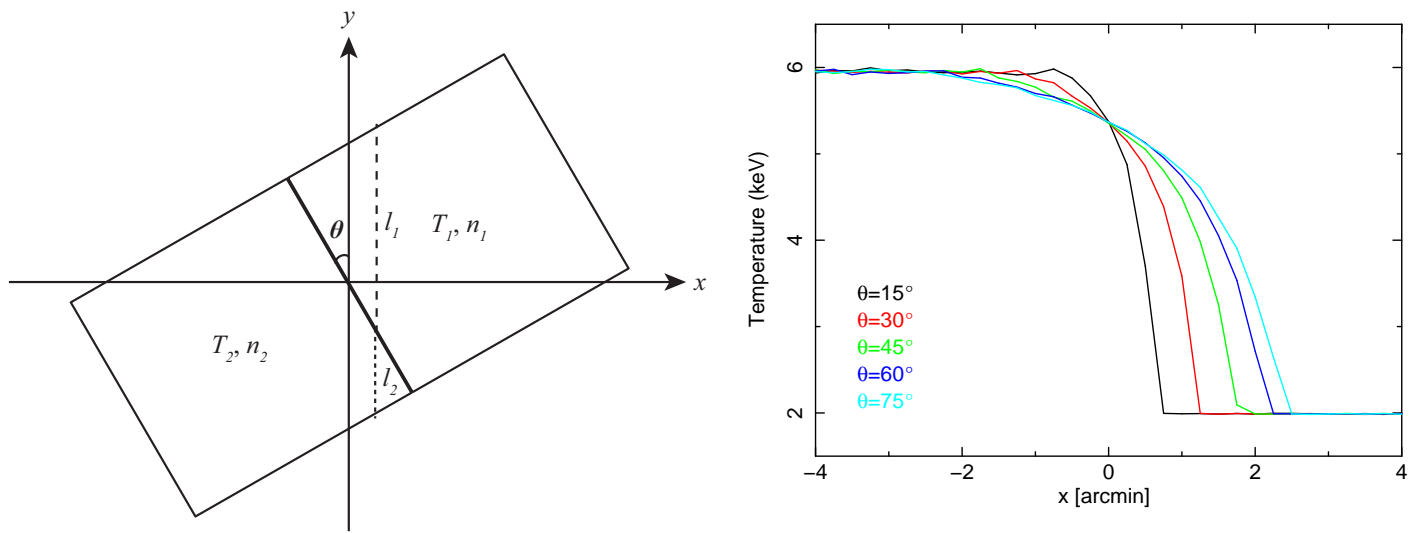}
%\end{center}
\end{minipage}
\begin{minipage}{.5\hsize}
%\begin{center}
\includegraphics[width=.7\hsize,angle=-90]{FIG17.ps}
%\end{center}
\end{minipage}
\end{tabular}
\caption{\label{fig:projection}
Left: ICM geometry assumed in calculating the projection effect (Sect. 4.4). The ICM temperature $T_i$, density $n_i$, and line-of-sight path length $l_i$ are indicated for the pre-shock ($i=1$) and post-shock ($i=2$) regions.  Right: Projected temperature profiles against the $x$-axis for the viewing angles of $\rm \theta = 15\degr (black), 30\degr (red), 45\degr (green), 60\degr (blue), and 75\degr$ (cyan).
}
\end{figure*}

\subsection{Possible scenarios for the discrepancy}
Even though Mach numbers from radio and X-ray observations are 
inconsistent with each other
, the expected Mach numbers in clusters are somewhat smaller (${\cal M}< 5$) than those in supernova remnants (${\cal M}> 1000$), which have an efficiency that is high enough to accelerate particles from thermal distributions to the $\sim$TeV regime~\cite[e.g., SN1006:][]{koyama95}. On the other hand, it is well known that the acceleration efficiency at weak shocks is too low to reproduce the observed radio brightness of radio relics with the  DSA mechanism \cite[e.g.,][]{kang12, vink14}.
To explain these puzzles, several possibilities have been proposed:
\begin{itemize}
\item projection effects that can lead to the underestimation of the Mach numbers from X-ray observations ~\citep{skillman13, hong15}
\item an underestimation of the post-shock temperature with electrons not reaching thermal equilibrium.  Such a phenomenon is observed in SNRs~\citep{vanadelsberg08, yamaguchi14,vink15}
\item Clumpiness and inhomogeneities in the ICM~\citep{nagai11, simionescu11}, which will lead to nonlinearity of the shock-acceleration efficiency~\citep{hoeft07}
\item pre-existing low-energy relativistic electrons and/or re-accelerated electrons,
resulting in a flat radio spectrum with a rather small temperature jump~\citep{markevitch05, kang12, pinzke13, kang15, store16}
\item a  nonuniform Mach number as a results of inhomogeneities in the ICM, which is expected in the periphery of the cluster~\citep{nagai11, simionescu11, mazzotta11}.
\item shock-drift accelerations, suggested from particle-in-cell simulations~\citep{guo14a, guo14b}
\item other mechanisms, for instance turbulence accelerations~\citep[e.g., ][]{fujita15, fujita16}
\end{itemize}

Based on our X-ray observational results,  we discuss some of the above possibilities.

{\it --ion-electron non-equilibrium after shock heating:}

Shocks not only work for particle acceleration, but also for heating. Immediately  after shock heating, an electron--ion  two-temperature structure has been predicted based on numerical simulations~\citep{takizawa99, akahori10}.  This could lead to an underestimation of the post-shock temperature because  it is hard to constrain the ion temperature with current X-ray satellites and instruments. 
  Therefore, the main observable is the electron temperature of the ICM, which will reach temperature equilibrium with the ion after the ion-electron relaxation time after shock heating.
With assumption of the energy transportation from ions to electrons through Coulomb collisions, the ion-electron relaxation time can be estimated as
\begin{equation}
t_{ie}=2\times 10^8 {\rm yr}(\frac{n_e}{10^{-3}~\rm cm^{-3}})^{-1}(\frac{T_e}{10^8~K})(\frac{\rm ln~ \Lambda }{40})
\end{equation}
\citep{takizawa99}. Here,  ln $\Lambda$ denotes the Coulomb logarithm~\citep{spitzer56}. 
From the APEC normalization, we estimated the electron density at the post-shock region as $n_{\rm e}\sim3\times10^{-4}~\rm cm^{-3}$, assuming a line-of-sight length of 1 Mpc.
Using
an electron density of $n_e=3\times10^{-4}~\rm cm^{-3}$ and $T_e=4.9~\rm keV$ at the post-shock region, we expect the ion-electron relaxation time to be $t_{i.e.}=0.3\times10^9\rm ~yr$. 
The speed of the post-shock material relative to the shock is $v_2=v_s/C\sim840 ~\rm km/s$. Thus, the region where the electron temperature is much lower than the ion temperature is about $d_{i.e.}=t_{i.e.}\times v_2=250$ kpc.
If there is no projection effect, then the estimated value is consistent with the integrated region that is used in the spectral analysis. This indicates that the ion-electron non-equilibration could be a possible cause of the discrepancy. We note that there are several claims that electrons can be more rapidly energized  (so-called instant equilibration) than by Coulomb collisions~\citep{markevitch10, yamaguchi14}.
However, it is difficult to investigate this phenomenon without better spectra than those from the  current X-ray spectrometer. The upcoming {\it Athena} satellite can shed new light on this problem.

{\it --Projection effects:}

Since it is difficult to accurately estimate the projection effects on the ${\cal M}_{\rm X}$ measurement, we calculate the temperature profiles to be obtained from X-ray spectroscopy under some simplified conditions. As shown in the left panel of Fig.~\ref{fig:projection}, we consider a pre-shock ($T_1 = 2$~keV) gas and a post-shock ($T_2 = 6$~keV) gas in $(5\arcmin)^3$ cubes whose density ratio is $n_2/n_1=3$ since the radio observation indicates a ${\cal M}\sim 3$ shock (Table~1). Assuming that the emission spectrum is given by a superposition of two APEC models and the emission measure of each model is proportional to $n_{i}^2 l_{i}$, 
we simulate the total XIS spectrum at a certain $x$  by the XSPEC fakeit command and fit it by a single-component APEC model.  
The right panel of Fig.~\ref{fig:projection} shows the resulting temperature profile against the $x$-axis for the viewing angles of $\theta=15\degr, ~30\degr, ~45\degr, ~60\degr, \rm and ~75\degr$. 
 Here the step-size was $\Delta x = 0\arcmin.2$, and 40 regions were simulated to produce the temperature profile for each viewing angle.
 This result indicates that the temperature of the pre-shock gas is easily overestimated as a consequence of the superposition of the hotter emission while the post-shock gas temperature is less affected. This result is consistent with the predictions by~\citet{skillman13} and \citet{hong15}. Thus the observed temperature ratio is likely to be reproduced when $\theta \gtrsim 30\degr$.

Next, we estimate the viewing angle based on the line-of-sight galaxy velocity distribution~\citep{yuan03}
 assuming that these galaxies in the infalling subcluster are move together with the shock front.
Because the brightest galaxy at the NE relic has a redshift of 0.0845, the peculiar velocity of the subcluster relative to the main cluster~\citep[0.0806:][]{struble99}
 is roughly estimated as $1180~{\rm km/s}$, yielding $\theta = \arctan{(v_{pec}/v_s)}= \arctan{(1180/1380)}=40.5\degr$. 
 This also suggests that the projection effect is not negligible in the present ${\cal M}_{\rm X}$ estimation. Our present simulation is based on the simplified conditions, however and is not accurate enough to reproduce the temperature observed at the outermost region where the projection effects are expected to be small. 
From the right panel of Fig.~\ref{fig:projection}, in case of  the view angle of $40.5\degr$, next to the shock ({\it r}=0.0\arcmin--2.0\arcmin) is expected to have a temperature higher than the outermost bin.
On the other hand, the observed temperature in the outermost region ($kT_{\rm outermost}=3.36^{+0.59}_{-0.38}$ keV) is consistent with that derived being located at immediately beyond the relic ($kT_{\rm just~beyond}=3.40^{+0.41}_{-0.15}$ keV). This indicates that either  the viewing angle is overestimated or the simulation is too simple. To address this problem, X-ray observations with better spatial and spectral resolutions are required.

{\it --Clumpiness and inhomogeneities in the ICM:}

\citet{nagai11} suggested that the ICM at the cluster peripheral regions is not a single phase, 
meaning that it cannnot be characterized by a single temperature and gas density, because of a clumpy accretion flow from large-scale structures. The degree of clumpiness of the ICM is charcterized by a clumping factor
$\displaystyle{
C\equiv{\langle \rho^2\rangle}/{\langle\rho\rangle^2}
}$. Here, {\it C}=1 represents a clump-free ICM.
 \citet{simionescu11} reported the possibility of a high clumpiness factor (10--20) at the virial radius based on {\it Suzaku} observations of the Perseus cluster. The inhomogeneities of the ICM generate a non-uniform Mach number, which could lead to the disagreement of  shock properties inferred from X-ray and radio observations because the shock acceleration efficiency 
 strongly nonlinearly depends on the Mach number~\citep{hoeft07}.
However, in the A2255, the NE relic is located well inside of the virial radius at $\sim$0.6 $r_{200}$. The simulations~\citep{nagai11} predict that the clumping factor at $r=0.5r_{200}$ is almost unity and  gradually increases to the virial radius ($C\sim2$).  Therefore, clumping is not likely to be the dominant source of the discrepancy.

\subsection{Magnetic fields at the NE relic}\label{sec:mag}
The diffuse radio emission is expected to be generated by synchrotron radiation of GeV energy electrons with a magnetic field in the ICM. These electrons scatter CMB photons via the inverse-Compton scattering. The nonthermal emission from clusters is a useful tool to investigate the magnetic field in the ICM.  Assuming that the same population of relativistic electrons radiates synchrotron emission and scatters off the CMB photons, the magnetic field in the ICM can be estimated by using the following equations for the synchrotron emission at the frequency $\nu_{\rm Syn}$ and the inverse-Compton emission at $\nu_{\rm IC}$ (Blumenthal \& Gould 1970)
\begin{eqnarray}
\frac{dW_{\rm Syn}}{d\nu_{\rm Syn} dt} & =& \frac{4\pi N_0 e^3 B^{(p+1)/2}}{m_e c^2}
\left(\frac{3e}{4\pi m_e c }\right)^{(p-1)/2}a(p)\nu_{\rm Syn}^{-(p-1)/2}, \label{eq4}\\
\frac{dW_{\rm IC}}{d\nu_{\rm IC} dt} &=& \frac{8\pi^2 r_0^2}{c^2}h^{-(p+3)/2}N_0(k T_{\rm CMB})^{(p+5)/2}F(p)\nu_{\rm IC}^{-(p-1)/2}, \label{eq5}
\end{eqnarray}
where $N_0$ and $p$ are the normalization and the power-law index of
the electron distribution, $r_0$ is the classical electron radius, $h$ is the Planck constant,
$T_{\rm CMB}$ is CMB temperature, and $T_{\rm CMB}=2.73(1+z)$~K.
The magnetic field in the cluster can be estimated by substituting observed flux densities of the synchrotron emission $S_{\rm Syn}$ and the inverse-Compton X-ray emission $S_{\rm IC}$ as well as their frequencies $\nu_{\rm Syn}$ and $\nu_{\rm IC}$ 
by the relation
 $S_{\rm Syn}/S_{\rm IC} = (dW_{\rm Syn}/d\nu_{\rm Syn} dt)/(dW_{\rm IC}/d\nu_{\rm IC} dt)$ (see also Ferrari et al. 2008; Ota et al. 2008; 2014).

Based on the X-ray spectral analysis assuming $\Gamma_{\rm relic} = 1.8$ for the nonthermal power-law component (Sect. \ref{sec:IC}), we derive the upper limit on the IC flux density as $S_{\rm IC}<0.067 $~mJy at 2~keV ($\nu_{\rm IC} = 2.9\times10^{18}$~Hz). Combining this limit with the radio flux density of the NE relic, $S_{\rm Syn}=117$~mJy at $\nu_{\rm Syn}=328$~MHz \citep{pizzo09}, 
we obtain the lower limit of the magnetic field to be
 $B > 0.0024~{\rm \mu G}$. Although the value depends on the assumptions, e.g.,
 such as spectral modeling and  area of the NE radio relic, the estimated lower limit is still much lower than the equipartition magnetic field ($B_{eq}=0.4~\mu$G) based on radio observations~\citep{pizzo09}.  A hard X-ray observation with higher sensitivity is needed to improve the accuracy and test the equipartition estimation~\citep[e.g., ][]{richard15}.

\section{Summary}\label{sec:summary}
Based on deep {\it Suzaku} observations of A2255 ({\it z} = 0.0806), we determined the radial distribution of the ICM temperature out to 0.9$r_{200}$. 
We found two temperature discontinuities at {\it r}=5\arcmin and 12\arcmin, whose locations coincide with the surface brightness drops observed in the {\it XMM-Newton} image. Thus these structures can be interpreted as shock fronts. We estimated their Mach numbers,  ${\cal M}_{\rm inner}\sim1.2$ and ${\cal M}_{\rm outer}\sim1.4$ using the Rankine-Hugoniot jump condition.
The Mach number of the inner shock is consistent with the previous XMM-Newton result \citep{sakelliou06}, 
but for the different azimuthal directions.
Thus the western shock structures reported by {\it XMM-Newton} and the northern structure detected by {\it Suzaku} might originate from the same episode of a subcluster infall. To examine this, 
we need a detailed investigation of the thermal properties
of the ICM in other directions, which is to be presented in our forthcoming paper.
The location of the second shock front coincides with that of the NE relic, indicating that the electrons in the NE relic have been accelerated by a merger shock. However, the Mach numbers derived from X-ray and radio observations, assuming basic DSA mechanism, are inconsistent with each other.  This indicates that  the simple DSA mechanism is not valid under some conditions, and therefore, more sophisticated mechanisms are required.

\section*{Acknowledgments}
We are grateful to the referee for useful comments that helped to improve this paper.
The authors thank the {\it Suzaku} team members for their support on the {\it Suzaku} project, R. Pizzo for providing the WSRT radio images. 
H.A and F.Z. acknowledge the support of NWO via a Veni grant.
Y.Y.Z. acknowledges support by the German BMWi through the
Verbundforschung under grant 50\,OR\,1506.
R.J.W. is supported by a Clay Fellowship awarded by the Harvard-Smithsonian Center for Astrophysics.
N.O. and M.T. acknowledge support by a Grant-in-Aid by KAKENHI No.25400231 and 26400218.
SRON is supported financially by NWO, the Netherlands Organization for Scientific Research. 

\bibliographystyle{aa}
\bibliography{A2255_aa}

\end{document}